\documentclass[twocolumn]{aastex63}

\setcitestyle{notesep={ },round,aysep={},yysep={;}}

\received{\today}
\revised{\today}
\accepted{\today}
\submitjournal{ApJL}

\shorttitle{Collimated fast jet in W43A}
\shortauthors{Tafoya et al.}

\begin{document}

\title{Shaping the envelope of the asymptotic giant branch star W43A with a collimated fast jet}

\correspondingauthor{Daniel Tafoya}
\email{daniel.tafoya@chalmers.se}

\author[0000-0002-2149-2660]{Daniel Tafoya}
\affiliation{Department of Space, Earth and Environment, Chalmers University of Technology, \\
	Onsala Space Observatory, 439~92 Onsala, Sweden}
\affiliation{Chile Observatory, National Astronomical Observatory of Japan, \\
	National Institutes of Natural Sciences, 2-21-1 Osawa, Mitaka, Tokyo, 181-8588, Japan}

\author[0000-0002-0880-0091]{Hiroshi Imai}
\affiliation{Center for General Education, Institute for Comprehensive Education, \\
	Kagoshima University, 1-21-30 Korimoto, Kagoshima 890-0065, Japan}
\affiliation{Amanogawa Galaxy Astronomy Research Center, Graduate School of Science and Engineering, \\
	Kagoshima University, 1-21-35 Korimoto, Kagoshima 890-0065, Japan}
\author[0000-0002-7065-542X]{Jos\'e F. G\'omez}
\affiliation{Instituto de Astrof\'isica de Andaluc\'ia, CSIC, Glorieta de la Astronom\'ia s/n, \\
	E-18008 Granada, Spain}

\author{Jun-ichi Nakashima}
\affiliation{Department of Astronomy, School of Physics and Astronomy, Sun Yat-sen University, \\
	Zhuhai 519082, China}

\author{Gabor Orosz}
\affiliation{School of Natural Sciences, University of Tasmania, Private Bag 37, \\
	Hobart, Tasmania 7001, Australia}
\affiliation{Xinjiang Astronomical Observatory, Chinese Academy of Sciences, 150 Science 1-Street, \\
	Urumqi, Xinjiang 830011, China}
\author{Bosco H. K. Yung}
\affiliation{Nicolaus Copernicus Astronomical Center, Rabia\'{n}ska 8, 87-100 \\
	Toru\'{n}, Poland}



\begin{abstract}

\noindent
One of the major puzzles in the study of stellar evolution is the formation process of bipolar and multi-polar 
planetary nebulae. There is growing consensus that collimated jets create cavities with dense walls in the 
slowly-expanding (10--20 ~km~s$^{-1}$) envelope ejected in previous evolutionary phases, leading to the 
observed morphologies. However, the launching of the jet and the way it interacts with the circumstellar material 
to create such asymmetric morphologies have remained poorly known. Here we present for the first time CO 
emission from the asymptotic giant branch star W43A that traces the whole stream of a jet, 
from the vicinity of its driving stellar system out to the regions where it shapes the circumstellar envelope. We found 
that the jet has a launch velocity of 175~km~s$^{-1}$ and decelerates to a velocity of 130~km~s$^{-1}$ as it interacts 
with circumstellar material. The continuum emission reveals a bipolar shell with a compact bright dot in the 
centre that pinpoints the location of the driving source of the jet. The kinematical ages of the jet and the bipolar 
shell are equal, $\tau$$\sim$60~years, indicating that they were created simultaneously, probably by a common 
underlying mechanism, and in an extremely short time.  These results provide key initial conditions for the 
theoretical models that aim to explain the formation of bipolar morphologies in the circumstellar envelopes of low 
and intermediate mass stars.

\end{abstract}



\section{Introduction} \label{sec:introduction}

In the past couple of decades there has been mounting observational evidence that, after the Asymptotic Giant Branch (AGB) 
phase, low and intermediate-mass ($M \lesssim 8M_{\odot}$) stars develop powerful highly collimated outflows that have a 
strong impact on the circumstellar envelope (CSE) that forms in previous mass-loss episodes. One important effect of the 
activity of such collimated outflows is sculpting and modifying the shape of the roughly spherical CSE \citep[e.g.][]{Sahai1998}. 
According to the increasingly popular model proposed by \cite{Sahai1998} to explain the formation of asymmetric planetary 
nebulae, the fast collimated outflows carve cavities within the spherical CSE that are eventually seen as bright lobes and 
bubble-like morphologies in the subsequent planetary nebula phase, when the star becomes hot enough to ionize the CSE.    
This process occurs in a very short time-scale ($\lesssim$1000~yr), during which  the star becomes enshrouded by gas and 
dust that render it invisible at optical wavelengths. As a result, it is difficult to find objects undergoing this ephemeral phase. 
Nonetheless, there is a particular group, containing 15 known sources in our Galaxy, of oxygen-rich post-AGB objects that 
exhibit high-velocity H$_{2}$O masers (v$_{\rm exp}$$\gtrsim$100~km~s$^{-1}$) tracing collimated structures and/or 
bow-shocks \citep{Imai2007}. These objects are called {\it water fountains} (WF) and they are thought to be undergoing 
the earliest manifestation of collimated mass-loss after the AGB phase \citep{Likkel1988}. Thus, the study of WF is of great 
importance to understand the evolution of low and intermediate-mass stars since they hold key information to understand the 
launching and collimation of jets in the late phases of these sources, as well as their interaction with the CSE.

W43A, located at a distance of 2.2~kpc from the Sun (see Appendix \ref{sec:W43A_distance}), is surrounded by a thick dusty 
envelope that renders it invisible at optical and near-infrared (IR) wavelengths \citep{Wilson1972}. The mid-IR images reveal an 
elongated structure (1$\rlap{.}^{\prime\prime}$2$\times$1$\rlap{.}^{\prime\prime}$6) oriented along a position angle 
(P.A.) $\sim$62$^{\circ}$ \citep{Lagadec2011}. The presence of hydroxyl (OH) masers with a periodic variability and the silicon 
monoxide (SiO) maser emission strongly suggest that W43A hosts an AGB star \citep{Herman1985,Imai2005}. Furthermore, the 
bipolar distribution of the SiO masers and the discovery of highly collimated water (H$_{2}$O) maser structures indicate that this 
source is undergoing the ephemeral transition in which the spherical CSE formed during the AGB phase develops a bipolar morphology 
\citep{Diamond1988,Imai2002,Imai2005,Vlemmings2006,Amiri2010,Chong2015}. The H$_{2}$O masers exhibit expansion velocities 
($V_{{\rm H}_{2}{\rm O}}$$\sim$150~km~s$^{-1}$) much higher than that of  the 1612 MHz OH masers
($V_{\rm OH1612}$$\sim$8~km~s$^{-1}$), which gives W43A the classification of a water fountain source \citep{Likkel1988,Imai2002}. 
The H$_{2}$O masers were originally 
found to be located mainly in two linear structures separated by 0$\rlap{.}^{\prime\prime}$6 on the plane of the sky, which 
corresponds to a de-projected separation of 1600~AU considering an inclination angle with respect to the plane of the sky of 
35$^{\circ}$ \citep{Imai2002,Chong2015}. The highly collimated motions and distribution of the H$_{2}$O masers suggested the 
presence of a jet with a kinematical age of only $\tau_{{\rm H}_{2}{\rm O}}$$\sim$50~yr \citep{Imai2002,Imai2005}. In addition, a 
magnetic field of $B$$\approx$200$\pm$75~mG was measured via the Zeeman effect of the H$_{2}$O masers \citep{Vlemmings2006}. 
Thus, it has been thought that the H$_{2}$O masers trace the regions where a magnetically collimated jet interacts with the CSE, 
creating low-density cavities that will lead to the formation of an asymmetric planetary nebula when the central star becomes hot 
enough to ionise the surrounding gas \citep{Imai2002,Vlemmings2006}. However, since the H$_{2}$O masers do not trace the jet in 
its entirety, the true extent of the jet, collimation mechanism, as well as the way it interacts with the ambient material have 
remained unclear \citep{Chong2015}.

\section{Observations} \label{sec:observations}

\begin{figure*}
	\begin{center} 
		\includegraphics[angle=0,scale=0.6]{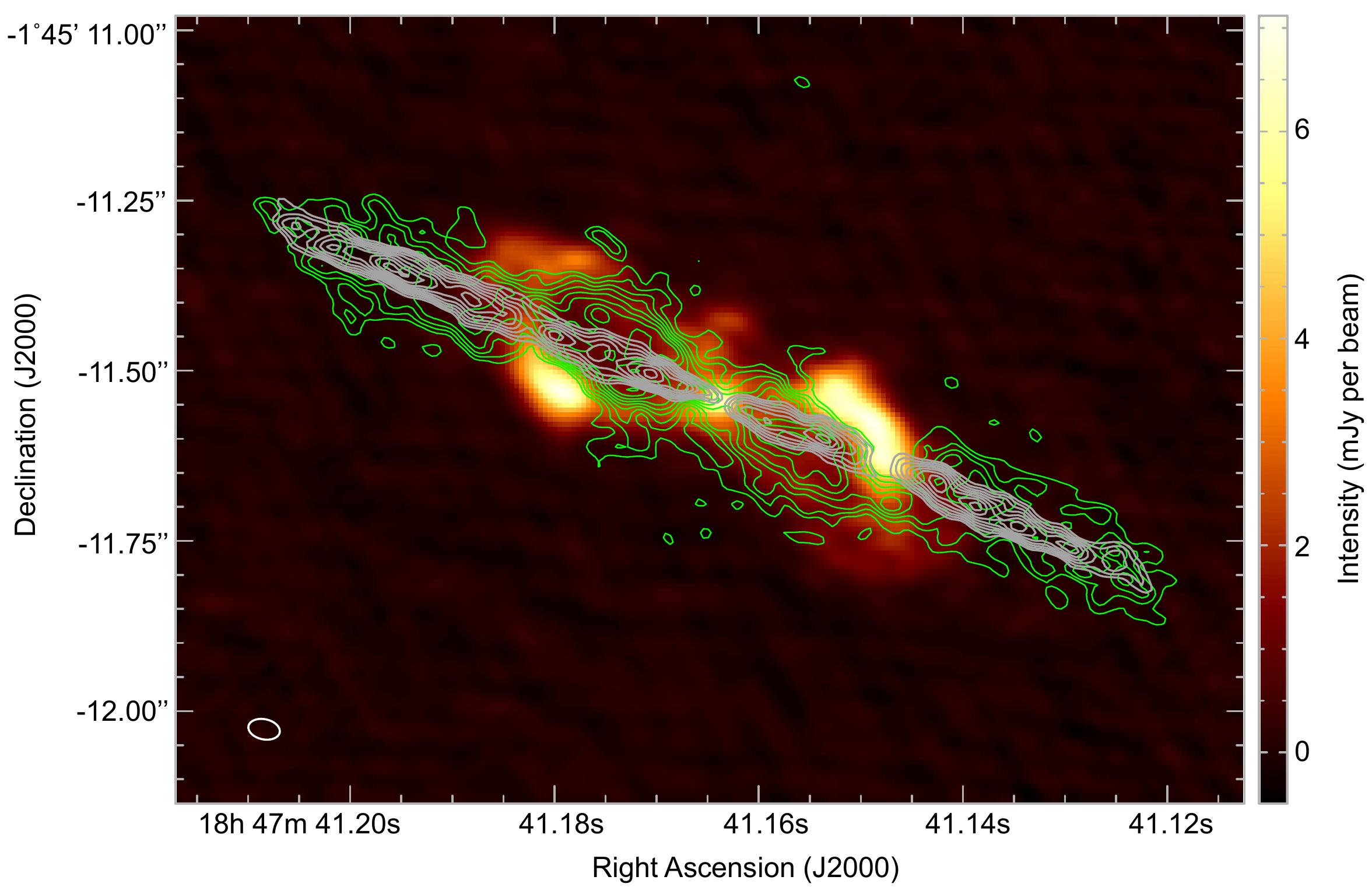}
		\caption{Continuum and CO($J$=2$\rightarrow$1) line emission from W43A. The green contours is CO($J$=2$\rightarrow$1) 
			emission integrated over the range of line-of-sight velocity offset $|$$V$$_{\rm offset}$(km~s$^{-1}$)$|$$<$75. The grey 
			contours is CO($J$=2$\rightarrow$1) emission integrated over the highest velocity range, 75$<$$|$$V$$_{\rm offset}$(km~s$^{-1}$)$|$$<$100. 
			The contours are drawn from 3-$\sigma$ on steps of 3-$\sigma$ ($\sigma$=15~mJy~beam$^{-1}$~km~s$^{-1}$). The ellipse at the 
			bottom-left corner represents the size of the synthesized beam of the CO observations,
			$\theta_{\rm beam}$=0$\rlap{.}^{\prime\prime}$048$\times$0$\rlap{.}^{\prime\prime}$031, 
			P.A.=74.8$^{\circ}$. The colour map is the continuum emission at $\sim$230~GHz and the corresponding synthesized beam is $\theta_{\rm beam}$=0$\rlap{.}^{\prime\prime}$047$\times$0$\rlap{.}^{\prime\prime}$030, P.A.=78.1$^{\circ}$.}\label{Fig1}
	\end{center} 
\end{figure*}

We used the Atacama Large Millimeter/submillimetre Array (ALMA) to observe with high-angular-resolution 
(0$\rlap{.}^{\prime\prime}$040=90~AU) the emission from the dust and molecular gas around W43A. 
The observations were carried out on 22 September 2017 using 42 antennas of the ALMA 12m array with Band 6 receivers 
($\sim$230~GHz) as part of the project 2016.1.00540.S (P.I. H. Imai). The maximum and minimum baseline lengths were 
12.1~km and 41.4~m, respectively. The corresponding angular resolution and maximum recoverable scale are 
0$\rlap{.}^{\prime\prime}$04 and 0$\rlap{.}^{\prime\prime}$6, respectively. 
The field of view of the observations was approximately 25$^{\prime\prime}$. The total time of observation on W43A was 23~min. 
The data were calibrated with the ALMA pipeline using J1924$-$2914 ($\sim$3.4~Jy) as a flux/bandpass 
calibrator and J1851+0035 ($\sim$260~mJy) as a gain calibrator. Images were created with CASA 5.4.0 using 
a Briggs weighting scheme with the robust parameter set to 0. The continuum emission was subtracted from the data 
cubes and channel maps with a spectral resolution of $\sim$1.3 km~s$^{-1}$ were created. The typical channel root-mean-square 
(rms) noise level is $\sim$2~mJy~beam$^{-1}$ in line-free channels. Among the detected spectral lines, 
the emission of the CO($J$=2$\rightarrow$1) and $p$-H$_{2}$S($J_{K_{\rm a},K_{\rm c}}$=2$_{20}$$\rightarrow$2$_{11}$) lines was 
used for the analysis presented in this work. The continuum image was created by averaging line-free channels from three 
234.375~MHz wide spectral windows giving a total bandwidth of 0.7~GHz. The central frequencies of the spectral windows 
used for creating the continuum are 230.282~GHz, 230.930~GHz and 231.866~GHz, respectively. The rms level in the continuum image is 
$\sim$110~$\mu$Jy~beam$^{-1}$.

\section{Results and Discussion} \label{sec:results_discussion}

\subsection{CO emission} \label{sec:CO_emission}

\begin{figure*}
	\begin{center} 
		\includegraphics[angle=0,scale=0.65]{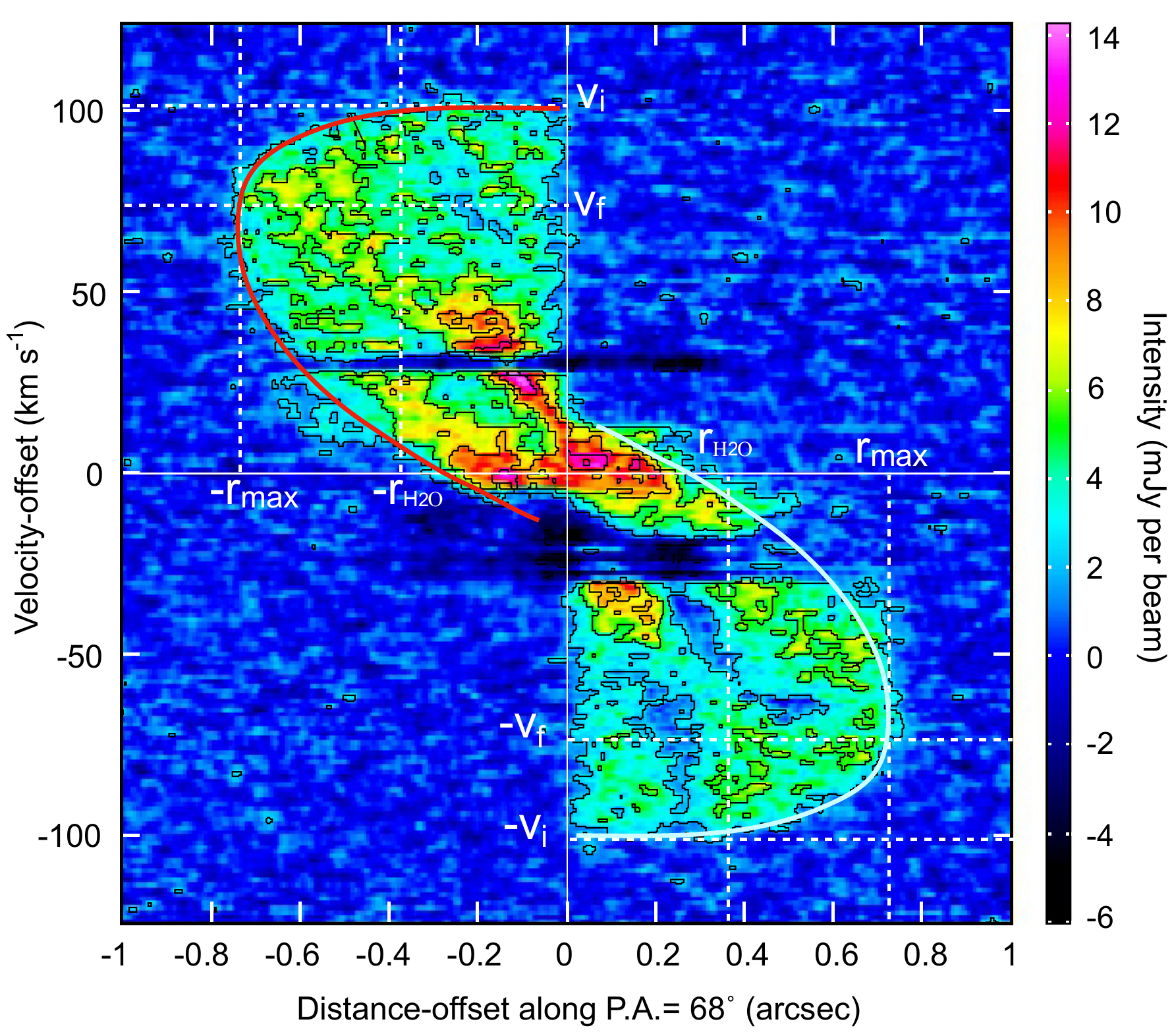}
		\caption{PV-diagram of the CO($J$=2$\rightarrow$1) line emission from W43A. The 
			PV-diagram was obtained using a slit along P.A.=68$^{\circ}$. The velocity offset is defined with 
			respect to the systemic velocity of the star $V_{\rm sys}$=34~km~s$^{-1}$. The horizontal white dashed lines 
			indicate the maximum and minimum line-of-sight velocity-offset of the emission that traces the jet (see main text). The locus 
			of the line-of-sight component of the velocity of the blue and red-shifted jet is indicated with light-blue and red solid lines, 
			respectively. The vertical dashed lines indicate the maximum extent of the CO($J$=2$\rightarrow$1) emission 
			along P.A.=68$^{\circ}$ and the approximate radius where the outer edge of the H$_{2}$O maser clusters are located, respectively. 
			The contours are drawn from 3-$\sigma$ on steps of 3-$\sigma$ ($\sigma$=1~mJy~beam$^{-1}$).}\label{Fig2} 
	\end{center}
\end{figure*}

The CO($J$=2$\rightarrow$1) emission spans a total velocity range of $\sim$200~km~s$^{-1}$ and is distributed over an elongated 
region (de-projected size $\sim$300$\times$3500~AU) whose major axis has a P.A.$\sim$68$^{\circ}$ (see Figures~\ref{Fig1} and ~\ref{FigA1}). 
From the channel map shown in Figure~\ref{FigA1} it can be seen that the emission with the highest velocity-offset\footnote{The 
velocity-offset, $V$$_{\rm offset}$, is defined with respect to the systemic velocity of W43A, $V_{\rm sys}$=34~km~s$^{-1}$, and corresponds 
to only the line-of-sight velocity component.}, $|$$V$$_{\rm offset}$(km~s$^{-1}$)$|$$\sim$100 (see also Figure~\ref{Fig2}), is the most collimated 
and lies closest to the star. For velocity-offsets 75$\lesssim$$|$$V$$_{\rm offset}$(km~s$^{-1}$)$|$$\lesssim$100, the emission exhibits a 
high degree of collimation and it reaches locations farther away from the star than for the highest velocity offsets, i.e. it exhibits a negative 
velocity gradient, $|$dv$|$/$|$dr$|$$<$0. Furthermore, for velocity-offsets 0$\lesssim$$|$$V$$_{\rm offset}$(km~s$^{-1}$)$|$$\lesssim$75 
the emission is less collimated and the velocity gradient inverts, i.e. the emission reaches locations farther away from the star with increasing 
velocity offset.

Thus, the CO($J$=2$\rightarrow$1) emission can be divided into two components characterised by their spatial distributions and velocity 
gradient. The component with negative velocity gradient has the highest velocity-offsets 75$\lesssim$$|$$V$$_{\rm offset}$(km~s$^{-1}$)$|$$\lesssim$100 
and it exhibits an extremely collimated morphology all the way from the vicinity of the central star ($\sim$90~AU) out to a de-projected distance of 
$\sim$1600~AU. The negative velocity gradient may indicate a deceleration of the emitting material as it moves away from the central system.
This component is shown with grey contours in Figure~\ref{Fig1}. The other component has lower velocity-offsets 
0$\lesssim$$|$$V$$_{\rm offset}$(km~s$^{-1}$)$|$$\lesssim$75 and is less collimated, albeit it has a bipolar morphology and is distributed around 
the high-velocity component, which is shown with green contours in Figure~\ref{Fig1}. The positive velocity gradient in this component may indicate 
entrainment of the circumstellar material by a fast jet.  The criteria to distinguish these kinematical components is further explained in 
\S\ref{sec:jet_entrained_material}.

\subsection{Continuum emission} \label{sec:continuum}

\begin{figure*}
	\begin{center} 
		\includegraphics[angle=0,scale=0.75]{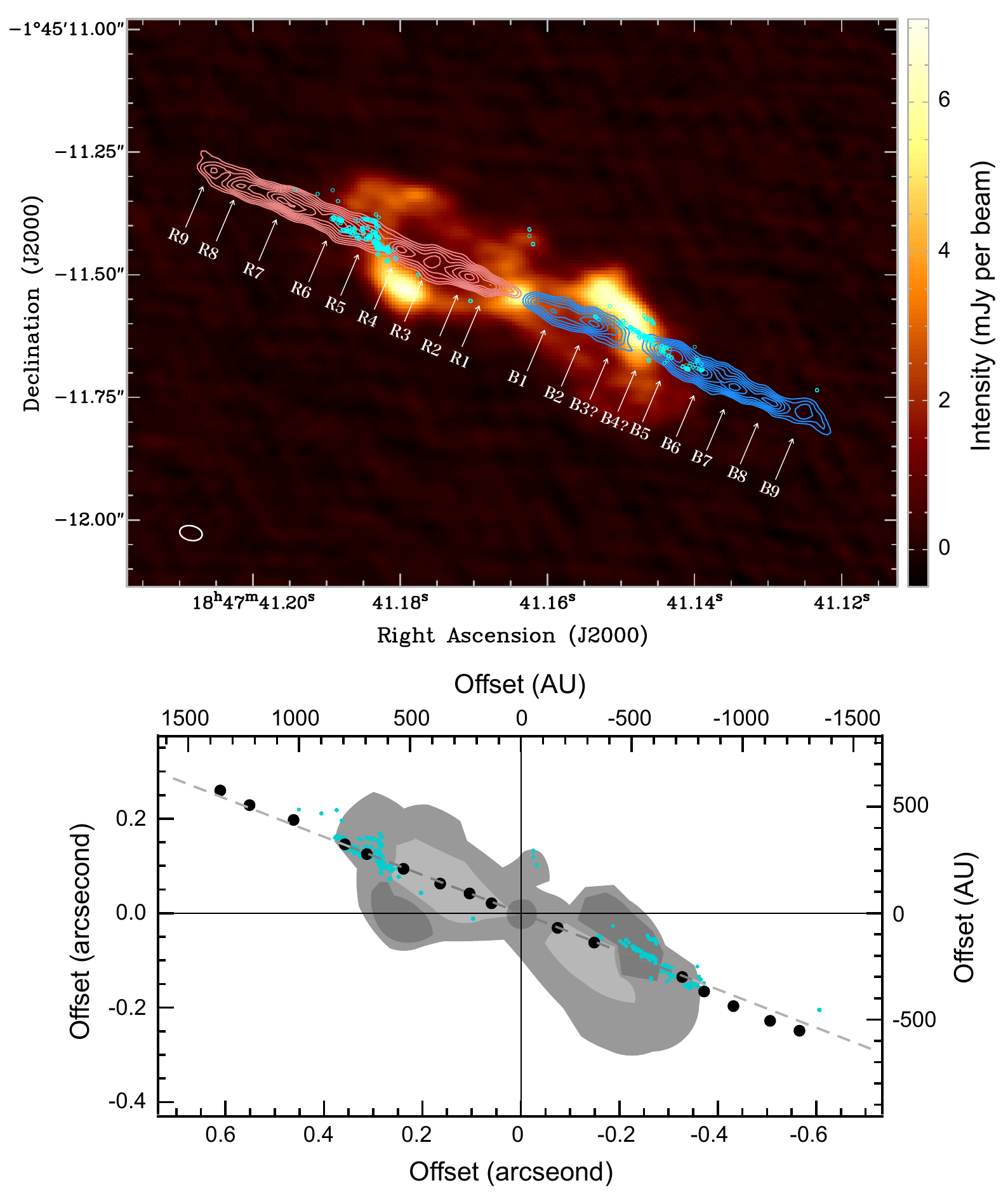}
		\caption{Upper panel: Continuum and CO($J$=2$\rightarrow$1) line emission from W43A. The blue and red 
			contours indicate blue and red-shifted CO($J$=2$\rightarrow$1) emission with velocity offsets $>$75~km~s$^{-1}$, tracing the jet.
			The contours are drawn from 3-$\sigma$ on steps of 3$\sigma$ ($\sigma$=15~mJy~km~s$^{-1}$). The colour map is the continuum 
			emission at $\sim$230~GHz. Lower panel: Diagram showing the location of the knots of the jet relative to the shell 
			traced by the continuum emission. The origin is the location of the emission peak of the central continuum 
			source:  (J2000) R.A.=18h 47m 41.164s, Dec.=$-$1$^{\circ}$45$^{\prime}$11$\rlap{.}^{\prime\prime}$54. The dashed line indicates a linear 
			fit to the clumps in the inner region of the bipolar lobes. The knots of the blue-shifted jet located 
			outside the lobes deviate from the linear fit. In both panels the H$_{2}$O masers found during 
			1994--2007 \citep{Chong2015}  and 2015 are indicated with dots in cyan colour (see Appendix~\ref{sec:VLBA_water_masers} for the details of the 
			H$_{2}$O masers observations).
		}\label{Fig3}
	\end{center} 
\end{figure*}

The  continuum emission traces a bipolar shell (colour map in Figure~\ref{Fig1}; see also Figure~\ref{Fig3}) that confines parts of the low-velocity 
CO($J$=2$\rightarrow$1) emission. The bipolar shell has a typical deconvolved width of 130~AU and it has two bright regions that exhibit point-symmetry 
with respect to the central source. From Figure~\ref{Fig1} it can be seen that the CO($J$=2$\rightarrow$1) emission is absorbed toward these bright regions, suggesting 
that they are optically thick regions. Therefore, the brightness temperature of the bright regions ($\approx 110$ K) is likely to be the actual temperature of the dust. 
Figure~\ref{Fig1} also reveals that the bright regions lie on the sides of the shell located closest to the main axis of the jet. This may be due to the 
passage of the jet close to the shell causing an increase in the density and/or temperature of the dust. The average density of the shell is 
estimated to be $n_{\rm H_{2}, Shell}$$\approx$5$\times$10$^{8}$~cm$^{-3}$ (see Appendix~\ref{sec:density_bipolar_shell} for the calculations), which favours the 
physical conditions to produce the observed H$_{2}$O masers. 

There is also a bright emission source at the centre of the nebula. A Gaussian fit to this source gives a deconvolved size of 90$\times$60~AU 
with a P.A. $\simeq 145^\circ$, which is nearly　perpendicular to the direction of the jet.

\subsection{A decelerating collimated jet and a bipolar outflow} \label{sec:jet_entrained_material}

Figure~\ref{Fig2} shows a PV-diagram of the CO($J$=2$\rightarrow$1) emission obtained with a slit along the major axis of the emitting 
region, P.A.=68$^{\circ}$. It is evident in this PV-diagram that the spatial spread of the CO($J$=2$\rightarrow$1) emission, delimited horizontally by the solid red 
and light-blue lines, increases as a function of the velocity-offset for 0$\lesssim$$|$$V$$_{\rm offset}$(km~s$^{-1}$)$|$$\lesssim$75 and decreases for 
75$\lesssim$$|$$V$$_{\rm offset}$(km~s$^{-1}$)$|$$\lesssim$100. Thus, this PV-diagram of W43A exhibits an S-like pattern that resembles the PV-diagram of the 
WF IRAS~16342$-$3814 \citep{Sahai2011,Tafoya2019}. \cite{Tafoya2019} interpreted the PV-diagram of IRAS~16342$-$3814 in terms of a jet-driven 
molecular outflow where the jet decelerates as it interacts with the surrounding material, transferring kinetic energy and momentum. The surrounding material 
and the jet produce the emission with positive and negative gradient, respectively, in the PV-diagram. Given the spatial distribution of the emission 
seen in Figure~\ref{Fig1} and the profile of the PV-diagram of Figure~\ref{Fig2}, 
it is evident that the spatio-kinematical configuration of W43A is similar to the one of IRAS~16342$-$3814. Following the analysis done by \cite{Tafoya2019}, 
the launch velocity of the jet can be obtained from the highest velocity-offset of 
the CO($J$=2$\rightarrow$1) emission, which in Figure~\ref{Fig2} is indicated as $|$$V_{\rm i}$$|$$\sim$100~km~s$^{-1}$. Considering the inclination of 
the source of 35$^{\circ}$ with respect to the plane of the sky, this velocity corresponds to a de-projected expansion velocity 
$U_{\rm i}$$\equiv$$|$$V_{\rm i}$$|$/sin~35$^{\circ}$=175~km~s$^{-1}$. The velocity-offset at which the velocity gradient changes from positive to 
negative is $|$$V$$_{\rm offset}$(km~s$^{-1}$)$|$$\approx$75. This value represents the velocity-offset at the tip of the jet, 
$V_{\rm f}$$\approx$75~km~s$^{-1}$, and its corresponding de-projected expansion velocity is 
$U_{\rm f}$$\equiv$$|$$V_{\rm f}$$|$/sin~35$^{\circ}$=130~km~s$^{-1}$. 

This shows that the material at the tip of the jet has decelerated during its journey from the 
launching region to its present location. This behaviour is expected if turbulent entrainment along the jet is a 
dominant process since, for such a case, the average velocity of the jet would decrease as momentum 
is transferred to the entrained material \citep{Chernin1994}. On the other hand, the overall velocity spread of the emission 
at a given position in the PV-diagram, delimited vertically by the solid red and light-blue lines, decreases with the distance from 
central system as an effect of the lateral expansion and turbulent motion of the gas \citep{Chernin1994, Balick2013}. From Figure~\ref{Fig2} it can be 
seen that beyond the radius of the outer edge of the H$_{2}$O masers clusters, $r$=$r_{{\rm H}_{2}{\rm O}}$, the maximum 
velocity-offset of the CO emission begins to drop significantly. Thus, it is likely that the largest deceleration 
occurs when the jet interacts with material of the shell, creating the physical conditions 
($T$$\approx$400~K; $n_{\rm H_{2}}$$\approx$10$^{9}$~cm$^{-3}$) that favour H$_{2}$O maser emission \citep{Yates1997}. 
This would explain the high concentration of H$_{2}$O masers in the interaction region. After the jet interacts with the shell, it continues 
streaming outwards and decelerates even further. 

The collimation factor of the CO structure that traces the jet, defined as the ratio between the length 
and the width, is very high ($\sim$20). Surprisingly, the degree of collimation of the jet is not affected after passing the 
region of interaction with the bipolar shell. The persistence of the high degree of collimation of the jet after the interaction 
with the shell is likely due to the pressure of the magnetic field, since it is expected to dominate the gas pressure by a 
factor of a few hundred in the interaction region \citep{Vlemmings2006}. 
Even though from these observations there is no clear evidence that the jet is precessing, as opposed to what was 
thought from the H$_{2}$O masers observations earlier \citep{Imai2002}, it is possible that there have been some slight 
variations in the P.A. of the jet that may explain the changes seen in the distribution of the H$_{2}$O masers over the past 
couple of decades \citep{Chong2015}.

\subsection{Time scale of the jet and bipolar shell} \label{sec:time_scales}

Considering that the velocity of the jet remains roughly 
constant from the launching point to the H$_{2}$O masers region, the kinematical time scale for the jet to reach 
this location is $\sim$30 years. In addition, assuming that the jet suffers a constant deceleration in the 
region beyond the location of the H$_{2}$O masers, the kinematical time scale from the H$_{2}$O maser region to the 
tip of the jet is $\sim$35~years. Thus, the total kinematical age of the jet is of just $\tau_{\rm jet}$=65~years. 

In addition to CO($J$=2$\rightarrow$1), the $p$-H$_{2}$S ($J_{K_{\rm a},K_{\rm c}}$=2$_{20}$$\rightarrow$2$_{11}$) line 
was also detected from the observations of this project. The H$_{2}$S emission is 
shown in Figure~\ref{FigA2} superimposed on the continuum emission. As it can be seen from Figure~\ref{FigA2}, the 
H$_{2}$S emission has a spatial distribution that is significantly different from that of the CO emission. Particularly, while the 
later one seems to be confined by the 
shell in the direction perpendicular to the outflow, the former one extends beyond the shell. Figure~\ref{FigA3} 
shows a PV-diagram of the H$_{2}$S
emission where it can be seen that it exhibits the characteristic morphology of a double-lobe structure, indicated with solid lines, 
similar to the morphology of the bipolar shell traced by the continuum emission. H$_{2}$S is known to trace shocked material 
\citep[e.g.][]{Holdship2016}, and in 
the case of W43A it is most likely tracing the shock produced by the expansion of the shell. Thus, we assume that the 
H$_{2}$S emission traces not only the spatial distribution but 
also the kinematics of the bipolar shell. The kinematical age of the bipolar shell can be obtained as 
$\tau_{\rm shell}$$\approx$$R_{\rm shell}$/$V_{\rm shell}$, 
where $V_{\rm shell}$ is the expansion velocity of the shell at any point located at a distance $R_{\rm shell}$ from 
the central source. The expansion velocity of the H$_{2}$S emission at the location of the H$_{2}$O masers 
($\sim$1100~AU) is $\approx$50~km~s$^{-1}$/sin($\theta_{\rm inc}$)=90~km~s$^{-1}$, where  $\theta_{\rm inc}$=35$^{\circ}$.)
The resulting timescale for the shell is $\tau_{\rm shell}$$\approx$60~years.

Thus, the kinematic ages of the jet and the bipolar shell, traced by 
the continuum emission, are basically equal, strongly suggesting that the former created the latter as 
it interacted with material in the vicinity of the driving source. The difference in size between the jet and 
the bipolar shell, despite their common kinematic age, can be explained in the following way. When the highly 
collimated jet was launched for the first time, around 65 years ago, it interacted with the dense material close to the central 
source creating bipolar shell. As circumstellar material was swept up, the expansion of the bipolar shell slowed down. 
Additionally, since the jet is launched with a velocity higher than the resulting velocity of the shell, it opens its way out through 
the shell, creating the physical conditions that favour the  H$_{2}$O 
maser emission in the region where the two components interact. Since the edges of the CO($J$=2$\rightarrow$1) emission are 
clearly defined and there is no evidence of collimated emission beyond the structures seen in Figure~\ref{Fig1} , it is likely that 
W43A is indeed undergoing the earliest phase of collimated mass-loss, as suggested previously \citep{Imai2002}. 
It is worth noting that recent numerical simulations presented by \cite{Balick2020}  predict 
morphologies very similar to that of W43A for certain combinations of initial conditions of the density, velocity 
and magnetic field. Therefore, these observational results are of utter importance to constrain such kind of numerical simulations.

\subsection{A binary system in W43A?} \label{sec:binary_system}

The CO($J$=2$\rightarrow$1) emission associated to the jet also exhibits a clumpy structure. The red-shifted jet has nine 
knots while the blue-shifted jet exhibits only seven knots (see Figure~\ref{Fig3}). The knots are roughly associated in pairs, although 
there are some variations in their separations from the central source. The missing knots in the blue-shifted jet coincide with the 
location of one of the continuum bright regions. Thus, it is likely that they are obscured by dense material in the dusty shell. 
The average knot spacing is 180~AU. Similar clumpy structures are commonly found in jets associated to young 
stellar objects \citep{Lee2007a,Hirano2006,Santiago-Garcia2009,Hirano2010} and they are interpreted as due to periodic events 
of increase of the velocity within the jets \citep{Santiago-Garcia2009,Hirano2010}. For the case of W43A, this could be explained 
in terms of a jet launching mechanism associated to accretion in a binary system that hosts an AGB star and a companion 
with eccentric orbits. When the components are in periastron the accretion disc is perturbed and the mass-loss rate 
increases \citep{Bonnell1992,Clarke1996,Davis2013}, leading to internal shocks in the jet that produce the observed clumpy 
structure \citep{Hirano2006,Santiago-Garcia2009}. Given the knot spacing and the speed of the jet, the time interval between 
them is around 5--7~years, which would be equal to the orbital period of the binary system. \\

\section{Final Remarks}

The observations presented in this work reveal with unprecedented detail the structure of a highly collimated jet and a bipolar 
bipolar morphology in the envelope of an AGB star and provide crucial information on the initial conditions of the CSE shaping. 
Particularly, the physical parameters derived from these observations will be essential to feed the theoretical models and numerical 
simulations that aim to explain the aspherical morphologies in the late stages of the evolution of low and intermediate mass stars 
\citep[e.g.][and references there in]{Balick2020}. Given the short time scale, $<$100~years, derived for this transition, it is imperative to investigate the 
other few ($\sim$15) water fountain sources \citep{Desmurs2012} that may exhibit different stages of this ephemeral phase.

\acknowledgments

This paper makes use of the following ALMA data: \newline ADS/JAO.ALMA\#2016.1.00540.S. 
ALMA is a partnership of ESO (representing its member states), NSF (USA) and NINS (Japan), together 
with NRC (Canada), MOST and ASIAA (Taiwan), and KASI (Republic of Korea), in cooperation with 
the Republic of Chile. The Joint ALMA Observatory is operated by ESO, AUI/NRAO and NAOJ. The 
VLBA is operated by the National Radio Astronomy Observatory (NRAO) under cooperative agreement 
by Associated Universities, Inc. JFG is partially supported by MINECO (Spain) grant AYA2017-84390-C2-R 
(co-funded by FEDER) and by the State Agency for Research of the Spanish MCIU 
through the ``Center of Excellence Severo Ochoa'' award for the Instituto de Astrof\'isica de Andaluc\'ia 
(SEV-2017-0709). HI and GO are supported by the MEXT KAKENHI program (16H02167). 
HI and JFG were supported by the Invitation Program for Foreign Researchers of the Japan Society 
for Promotion of Science (JSPS grant S14128). GO was supported by the Australian Research Council 
Discovery project DP180101061 of the Australian government, and the grants of CAS LCWR 2018-XBQNXZ-B-021 
and National Key R\&D Program 2018YFA0404602 of China. DT was supported by the ERC consolidator grant 614264. 
The authors are grateful to Bruce Balick for careful reading of the paper and valuable suggestions and comments.

\appendix

\section{Distance to W43A.} \label{sec:W43A_distance}
In previous works, W43A has been assumed to be located at a distance of $D$=2.6~kpc from the Sun. This 
value was calculated assuming the source is in a circular orbit around the Galaxy with an angular velocity given 
by a Galactic rotation curve derived in the 1960s \citep{Schmidt1965,Bowers1980}. In this work, the  
kinematical distance to W43A is updated using the Revised Kinematic Distance Calculator (2014) using Galactic 
parameters from model A5 in \cite{Reid2014}. The calculator gives near and far kinematical distances of 
2.22$^{+0.38}_{-0.40}$~kpc and 11.90$^{+0.38}_{-0.36}$~kpc, respectively. We adopt the near kinematical 
distance since it is compatible with the result of the Bayesian Distance Calculator \citep{Reid2016}. 

\section{Channel map of the CO($J$=2$\rightarrow$1) emission in W43A.} \label{sec:CO_channel_map}

\begin{figure*}[!h]
	\begin{center} 
		\includegraphics[angle=0,scale=0.7]{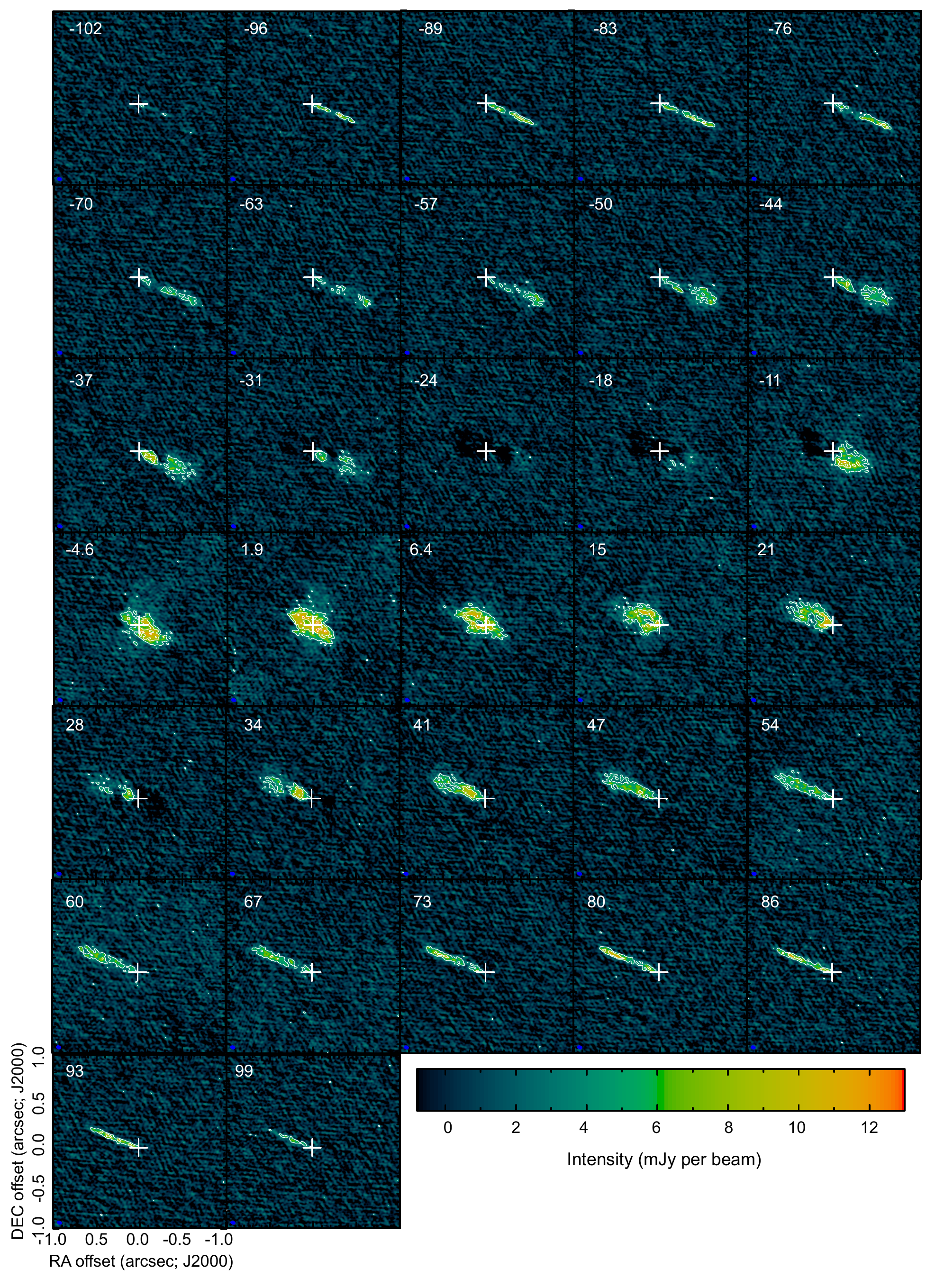}
		\caption{Velocity channel map of the CO($J$=2$\rightarrow$1) emission from W43A. The contours indicate 
			5, 10 and 15 times the rms noise level of the maps, 8$\times$10$^{-4}$~Jy~beam$^{-1}$. The numbers in the 
			upper left corner correspond to the velocity-offset with respect to the systemic velocity (34~km~s$^{-1}$), $V$$_{\rm offset}$, 
			of the channel map. The vertical colour wedge is in units of Jy beam$^{-1}$. The white cross is located at the position 
			of the central bright dot that pinpoints the location of the star: 
			(J2000) R.A.=18h 47m 41.164s, Dec.=$-$1$^{\circ}$45$^{\prime}$11$\rlap{.}^{\prime\prime}$54. The blue ellipse 
			located at the lower left corner indicates the synthesised beam, 
			$\theta_{\rm FWHM}$=0$\rlap{.}^{\prime\prime}$048$\times$0$\rlap{.}^{\prime\prime}$031, P.A.=75$^{\circ}$.}\label{FigA1} 
	\end{center} 
\end{figure*}

Figure~\ref{FigA1} shows a velocity channel map of the CO($J$=2$\rightarrow$1) emission from W43A. 

\restartappendixnumbering

\section{Velocity-integrated and PV-diagram of the H$_{2}$S line emission in W43A.} \label{sec:CO_channel_map}

\begin{figure}
	\begin{center} 
		\includegraphics[angle=0,scale=0.7]{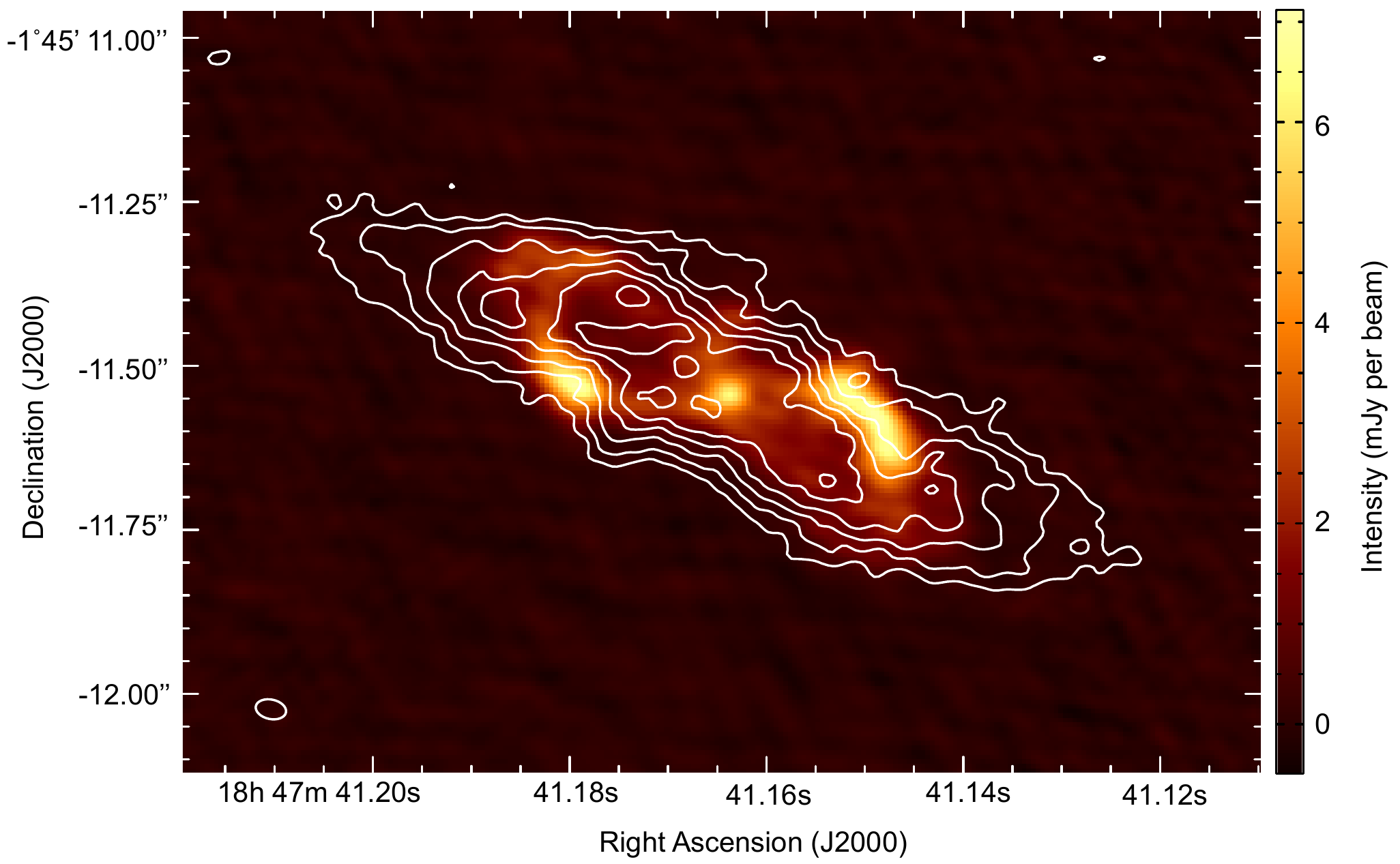}
		\caption{Continuum and $p$-H$_{2}$S($J_{K_{\rm a},K_{\rm c}}$=2$_{20}$$\rightarrow$2$_{11}$) line emission from W43A. 
			The contours represent H$_{2}$S emission integrated over 
			the velocity range -100$<$$V$$_{\rm offset}$(km~s$^{-1}$)$<$+100. The contours are (100, 200, 300, 400, 500, 600) 
			mJy~beam$^{-1}$~km~s$^{-1}$ ($\sigma$=35~mJy~beam$^{-1}$~km~s$^{-1}$). The ellipse at the bottom-left corner 
			represents the size of the synthesized beam $\theta_{\rm beam}$=0$\rlap{.}^{\prime\prime}$071$\times$0$\rlap{.}^{\prime\prime}$048, 
			P.A.=86.3$^{\circ}$. The colour map is the continuum emission at $\sim$230~GHz.}\label{FigA2} 
	\end{center} 
\end{figure}

\begin{figure}
	\begin{center} 
		\includegraphics[angle=0,scale=0.75]{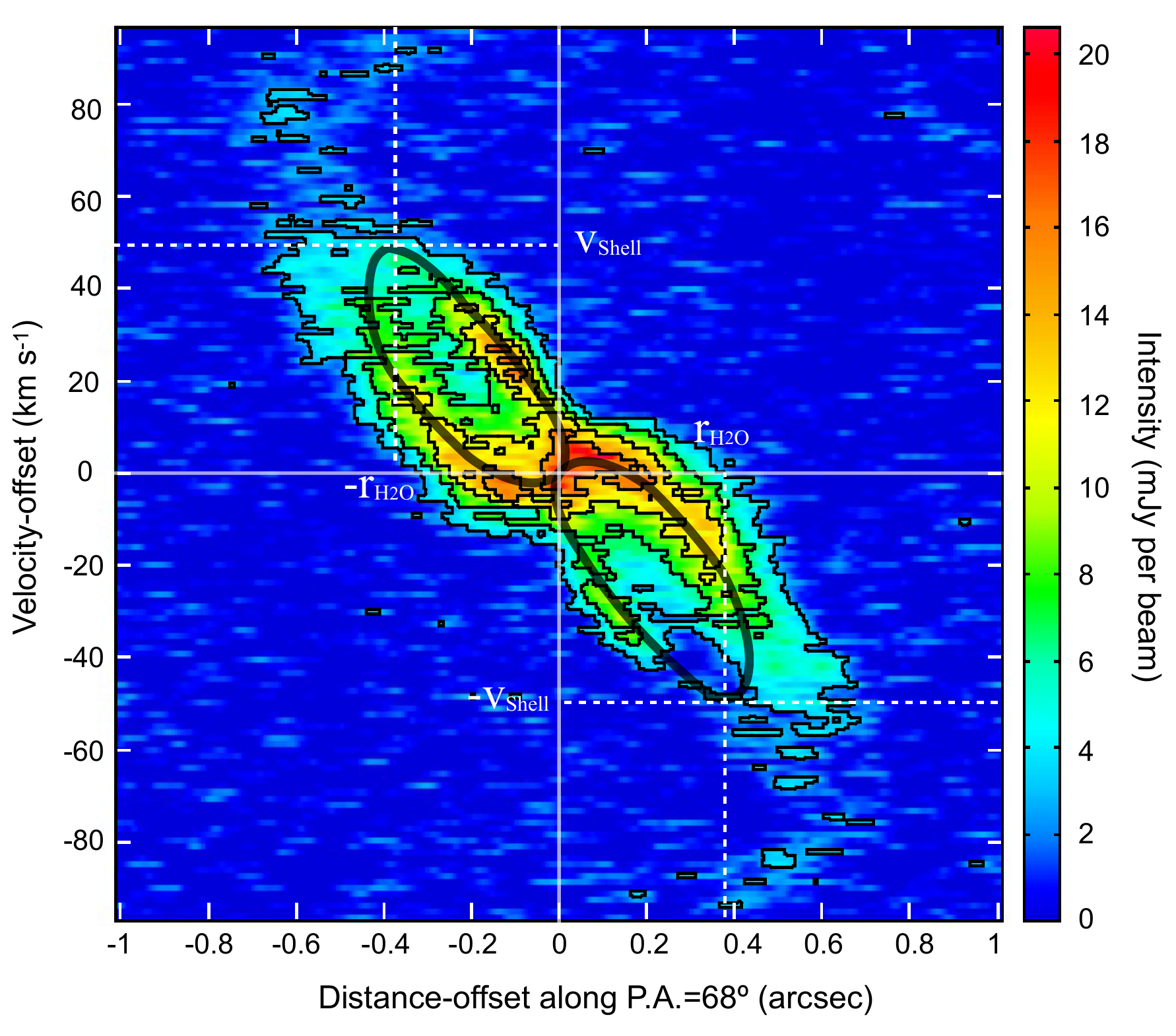}
		\caption{PV-diagram of the $p$-H$_{2}$S($J_{K_{\rm a},K_{\rm c}}$=2$_{20}$$\rightarrow$2$_{11}$) line emission from W43A. 
			The H$_{2}$S emission exhibits the characteristic morphology of a double-lobe structure, indicated with a solid black line, 
			similarly to the shell delineated by the continuum emission. The velocity offset is 
			defined with respect to the systemic velocity of the star $V_{\rm sys}$=34~km~s$^{-1}$. The vertical dashed lines indicate the 
			radius where the H$_{2}$O masers are located. The horizontal white dashed lines indicate the maximum velocity-offset along the 
			line-of-sight of the shell traced by the $p$-H$_{2}$S($J_{K_{\rm a},K_{\rm c}}$=2$_{20}$$\rightarrow$2$_{11}$) emission. The 
			contours are from 4-$\sigma$ on steps of 4-$\sigma$ ($\sigma$=1~mJy~beam$^{-1}$)}\label{FigA3} 
	\end{center} 
\end{figure}

The contours in Figure~\ref{FigA2} represent a velocity-integrated map of the $p$-H$_{2}$S($J_{K_{\rm a},K_{\rm c}}$=2$_{20}$$\rightarrow$2$_{11}$) 
emission in W43A superimposed on the continuum emission (colour map) at $\sim$230~GHz. Figure~\ref{FigA3} shows PV-diagram of the H$_{2}$S 
emission obtained using a slit along P.A.=68$^{\circ}$.

\section{Volume density of the molecular gas in the bipolar shell.}  \label{sec:density_bipolar_shell}
The mass of the gas in the bipolar shell can be estimated assuming optically thin emission and isothermal conditions for the dust. 
It should be pointed out that although the dust emission is optically thick toward the bright regions, as discussed in the main text, 
it is reasonable to assume optically thin emission from the rest of the shell. Having this in mind, the mass of the dust can obtained 
from the following relation \citep{Hildebrand1983,Gall2011}:
\begin{equation}\label{Eq:3}
M_{\rm d}=\frac{D^{2}\,S_{\nu}}{\kappa_{\nu}B_{\nu}(T_{\rm d})}, 
\end{equation}
where $D$ is the distance to the source, $S_{\nu}$ is the flux density of the continuum emission, $\kappa_{\nu}$ is 
the dust absorption coefficient and $B_{\nu}(T_{\rm d})$ is the Planck function evaluated at the temperature of the 
dust, $T_{\rm d}$. The flux density of the continuum emission that traces the bipolar shell is 267$\pm$25~mJy. 
Considering an average dust temperature $T_{\rm d}$=100~K, obtained from the brightness temperature of the 
optically thick bright regions; a dust absorption coefficient $\kappa_{\nu}$=1.8~cm$^{2}$~g$^{-1}$ for small silicate grains 
with coating of amorphous carbon \citep{Ossenkopf1994}, and a dust-to-gas ratio value of 100, the mass of the 
bipolar shell is 2.0$\pm$0.2 solar masses. The volume of the bipolar shell is obtained from the size of the bipolar 
lobes of molecular gas confined by the shell and from the thickness of the shell. The resulting average density of 
the bipolar shell is $n_{\rm H_{2}, Shell}$=5.1$\pm$0.5$\times$10$^{8}$~cm$^{-3}$. For silicate grains without 
coating the dust absorption coefficient is $\kappa_{\nu}$$\approx$0.15~cm$^{2}$~g$^{-1}$ \citep{Ossenkopf1994}, 
resulting in values of the bipolar shell mass and density a factor of 10 larger, which are unrealistically high given 
that the initial mass of the star must be $M$$\lesssim$8$M_{\odot}$.

\section{VLBA observations of the H$_{2}$O masers.} \label{sec:VLBA_water_masers}

The results of all the VLBA observations presented in this paper were summarized by \cite{Chong2015}. 
One exception is the one based on the latest observation on 19 January 2014 (project code BI41A). This observation 
and its data reduction were conducted similarly to those of other 
observations. The detected H$_{2}$O masers include a maser feature located close to the tip of the 
blue-shifted jet,  and give the largest extent of the maser distribution. This indicates that the H$_{2}$O maser 
region has grown along the jet at a rate roughly consistent with the proper motions of the individual maser 
features ($\sim$9~milliarcse\-conds~yr$^{-1}$ in one side of the bipolar jet). The coordinates of the fringe-phase and 
position reference maser spot at $V_{\rm LSR}$=126.1~km~s$^{-1}$ were determined to be: 
(J2000) R.A.=18h 47m 41.1821s, Dec.=$-$01$^{\circ}$45$^{\prime}$11$\rlap{.}^{\prime\prime}$396 , with an 
uncertainty better than 1~milliarcsecond that may be dominated by the uncertainty in the position of the 
phase-reference source J183323.9$-$032331. The superposition of the H$_{2}$O maser maps taken from 
the different observation sessions was made by the spatio-kinematical fitting to the bipolar jet as 
described by \cite{Chong2015}. 


\bibliography{bibliography_tafoya}{}
\bibliographystyle{aasjournal}



\end{document}